UNIVERSITY SINGIDUNUM

FACULTY OF TECHNICAL SCIENCES

# Climb Against Time – Self-perspective through a psychological game
- research report -

Author:                                                         Co-Author:

*Stefan Cliff*                                        Prof. Dr *Mlađan Jovanović*

Belgrade, *2023*.

# Table of contents

# Introduction

With the rapid development of technology and its place in our lives, so too has the idea of needing to *grow up* faster, do more, be more and more as we are exposed to so many of our *betters* billboarding their successes and achievements that very often we can experience burnout, depression, feeling of inadequacy and worse. All because we cannot keep up with their tempos in life, and in this chaos, we often lose the very important fact and truth, our life should be lived at the tempo that fits our actual wants, our capabilities and opportunities.

This problem of feeling disappointed, angry, or lost all due to you not being able to match those you hear about has been going on for a very long time. And with access to more and more information nowadays, so too comes to access to such not always intentionally harmful messages or *facts.* Looking outward is a part of the human psyche, but too much can lead to looking inward and only seeing failure, doubt and more. We are told that if we don't do X, Y or Z then we will fail, it won't work and if we can't even do that *simple* thing, then we won't work and never will till we fix ourselves by throwing ourselves into another metaphorical circle of different kinds of success. Often this can be seen in a vainer lifestyle that seems to promote health and success, but feeds a cycle of diminishing self-respect, care, and appreciation for the achievements you accomplished.

In recent years, since the mid 2010's, video games have entered the mainstream even more than before as a media platform that provides a more interactive experience than others like it. Where the players actions have consequences, outcomes both *good* and *bad*, and the experience of the player is highly linked to their capabilities. Based on the type of video game, be it single-player or multiplayer, often the solution to the problem the player is facing (i.e., a level that is very difficult, a specific enemy, etc.) will vary. For single-player games, the number of solutions is grander and the way to implement them is easier to monitor, test and incorporate into the game itself. For example, lowering the health of the enemies the player is facing, increasing the players health or how fast or high they can jump, giving them more lives and retries. And for a long while, single-player games were the main pinnacle of the industry, though difficult to create and polish and often led to longer development time as well as a high chance of not getting a high return in sales. With that, so came the next wave of gaming. A higher focus on skill-based, fast-paced, repeatable multiplayer games whose matches were created via a series of algorithms to compare statistics and statistics alone of each player all of which culminates into one, easy to compute and compare value. This value often has its own unique name given to it by the developers, but the most common one is Matchmaking Rating (MMR). Think of it as a secret sauce recipe that assigns you a numerical value based on your performance (Dana Douglas, p.1). How MMR is determined varies from system to system, some ask players to complete a certain number of games to calculate the statistics based on their performance and assign it to the player. Later, as the player continues to participate in more *matches* or *rounds* their MMR might change, be it for the better or worse. In her blogpost, Dana walks us through the most common ways MMR is calculated in competitive games, the ranges of the values which are usually from 0 (zero) all the way to 10 000 (ten thousand) with the average gamer falling around the 3000 (three thousand) mark.

Often each game platform has a way to let the player know where they fall without revealing the exact number of their MMR. I will be using the popular multiplayer online battle arena developed and published by Riot Games, League of Legends (LoL), as an example what this presentation and, ultimately, separation looks like. In LoL or League, each player is assigned to one of two teams of six. Depending on the game mode, the map or arena they fight in will or will not have multiple *lanes,* each of which will have a constant stream of computer controlled Non-Playable Characters (NPCs) called *minions.* The aim of any of the following game modes is the same, level up, obtain gold, defeat the other team, and destroy their *Nexus* which is at the heart of the enemy's base, all whilst defending their own *Nexus* from being destroyed as well. The players have multiple game modes to choose from and they are:

1. *Summoners Rift*, which has its own sub specific game modes one of which is Ranked.
2. *All Random All Madness* (ARAM), which does not have the same sub genres as Summoners Rift.

After selecting the game mode and a game being created based on your MMR, there is a champion picking phase. With over 140 (one hundred and forty) champions available (Riot Games 2023) in the game, the complexity and tug-of-war like strategy starts as soon as the game begins. With some champions being weaker or stronger than others. After playing the necessary number of Ranked games on Summoners Rift, the player is bestowed a rank. The ranks available are, in ascending order:

1. Iron
2. Bronze
3. Silver
4. Gold
5. Platinum
6. Emerald
7. Diamond
8. Master
9. Grandmaster
10. Challenger

With each rank having its own division, going from 4 (four) to 1 (one). There are further specifics to do with the ranking system that Riot has developed, but we believe that they aren't necessary to mention.

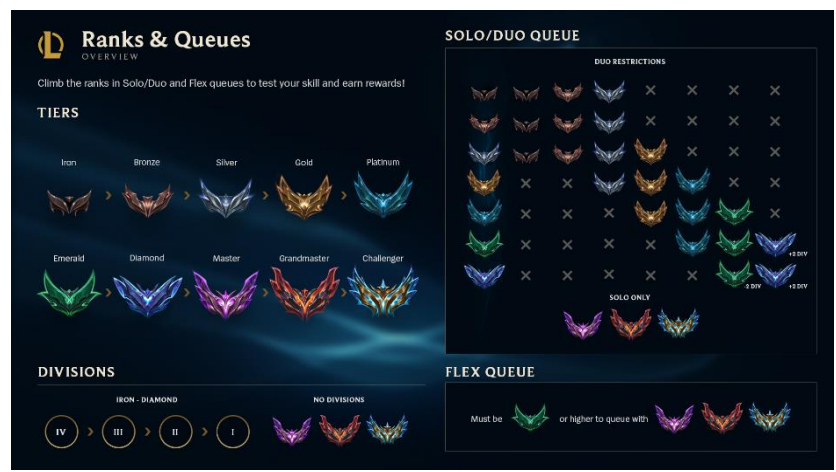

*Figure 1 – Ranked Tiers, Divisions, and Queues (Riot Games 2023.)*

With this ranking system in play, it often leads to a better and more fair experience for the average player. However, one core reason that many people enjoy playing video games is to escape or have fun. For some, especially those who are on the younger side, it can influence their perception of themselves and their abilities. Be it enjoying a rush of dopamine after a hard-fought victory or anger and frustration following a crushing defeat, whilst the enviroment may be virtual the emotions and impact is very much so real. These feelings do leak into our day to day lives, inspiring or numbing in cases. And if games are truly meant to be a means of escape and fun, what kind of effect can that have on us outside of the game if we end up being unable to be skilful enough to get better whilst being bombarded with the everyday new onslaught of people now saying how the game is meant to be played otherwise you will never win. In the day and age where technology is all around us, in our pockets, on our wrists in some cases, the access to competitive games is spreading at almost the same speed and availability. The nature of these competitive games usually demands more and more time, which the average working adult often lacks, and often means that a larger portion of these competitive multiplayer games are being played by a younger demographic. And according to the findings of Karol Severin (2022.) who wrote a blog for Midia Research, the gap between single-player and multiplayer preference numbers grows more with the age of the gamer. He further writes that this might be linked to the older generations growing up on primarily local based games that had no means of connecting to other devices and likely the disconnect and slowdown of reaction speeds for the average player.

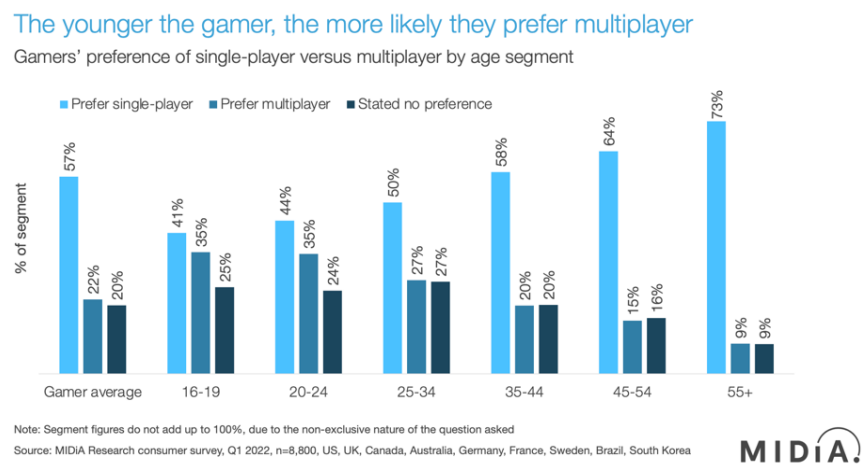

*Figure 2 - The Younger the gamer, the more likely the prefer multiplayer (Karol Severin 2022.)*

Further-on, in more recent years, since approximately the end of 2020, there has been a great rise in available tools for the public to create on their own time. Be it AI platforms to help give book recommendations, for example, or developer companies creating applications where users can download and being creating their own games, art, cinematic videos, render their own models, create music and so much more. This wave of available tools to create has also led to a rise in indie (short for independent) developers or teams of indie developers creating their own games for the public. "When it comes to gaming, *AAA* means that the has been or will be released by a major publisher (such as EA, Ubisoft, Sony, Microsoft, Blizzard, etc.) and typically indicates that is has undergone extensive, meticulous development and had a massive budget." (Adzo 2023.). This style of creating games is often quite different in scope and financing as opposed to Triple-A, but their rising popularity is not something that can be ignored. As VG Insights say that on Steam, on one of if not the most popular PC gaming platform, just over 95% of all the games available for purchasing are indie games however they lead to 40% of all the sales made on the platform (VG Insight 2021.).

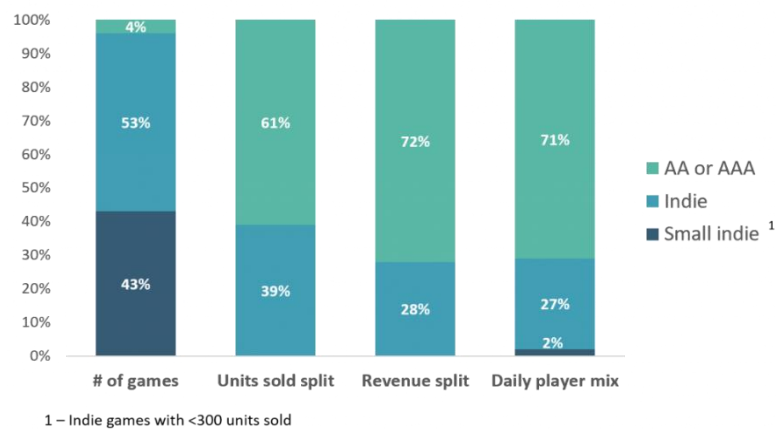

*Figure 3 - Distribution of Games Based on Publisher Type (VG Insight 2021.)*

With all that said, with the increase popularity of both buying and creating games, more and more personal stories, talented teams and individuals, unique takes and ideas are being tried and often for the games that do succeed, there is financial gain but more impactful is communities built around the message and/or its execution. These communities usually reside on social media, such as Reddit, X, Tumblr, or on their own community pages made by the developers to directly interact with their players. But this is true often even for games that don't gain much financial success, they gain a certain cult following, especially if the topics of the game are either obscure or whose workings revolve around mental health issues, trauma survival, loss, or just in general very human and emotional topics. One such game gained a very large following and financial success after its release is Celeste, developed by Maddy Makes Games, later rebranded to Extremely OK Games. "Celeste has gained a massive fanbase thanks to its realistic depiction of dealing with mental health issues and questioning your own identity with a heartfelt but realistic eye". (Willa Rowe 2023.). It was one of the largest hits of that year and it inspired so many to develop and work on their own games more than before. It took players on a journey and told a story of self-discovery both through other character interactions as well as various metaphors placed all around the hundreds of levels. It inspired me to take a greater look at how narrative and world design can both intwine and create something that feels different yet familiar.

# 1 Unity for Game Development

Unity is commercially available game engine with its own integrated development enviroment (IDE) which was developed and introduced by Unity Technologies back in 2005 with the intent for the product to be used for creating a wide variety of interactive media (video games, 3D renderings and animations, etc.). At first it was created as a Mac OS X Game engine, but since then has been gradually adding support for a wide variety of platforms such as: desktops, mobile, console and virtual reality. With platform neutral input management systems in place, it leads to generally a less volatile development lifecycle and a greater starting point for multi-platform development even if they have different input styles, for example keyboard & mouse and joystick controller.

To begin creating using the Unity Engine, you must first go to the Unity website https://unity.com, and create a Unity account. After which you will download their engine version and project manager application called Unity Hub.

## 1.1 A Quick Guide to Unity Hub

After downloading the application, you will be asked to login to your Unity account and select a version of their engine to download. When the download finishes, you'll be taken to the main part of the application, which looks like so:

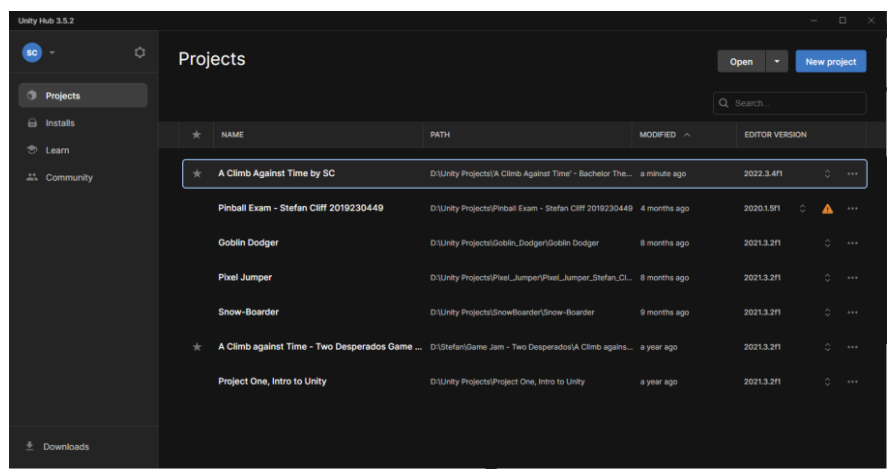

*Figure 4 - Unity Hub Main Window*

From here you'll be able to select any of you currently opened projects if you have any or create a new one. Unity offers many template setups for the genre or platform setup for the project you wish to create. After naming, selecting a template and pressing create you'll be taken to the Unity Editor.

**1.2 The Unity Editor**

Before talking about the workflow of the Unity Editor, first I must explain what one of the cornerstones of development inside Unity is. That cornerstone being GameObjects. "GameObjects are the fundamental objects in Unity that represent characters, props, and scenery. They do not accomplish much in themselves, but they act as containers for Components, which implement the functionality." (Unity Documentation 2023).

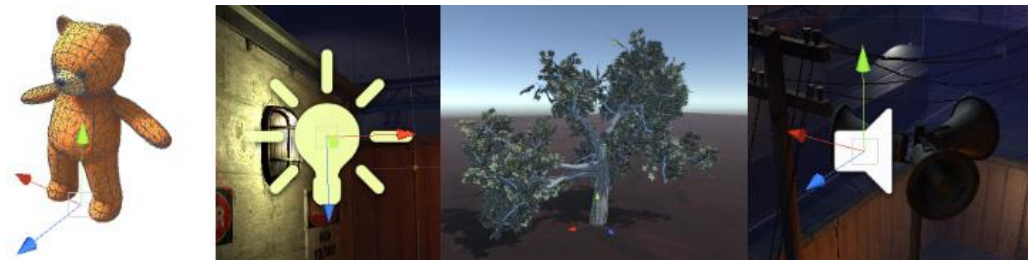

*Figure 5 - Four different types of GameObjects: an animated character, a light, a tree, and an audio source (Unity Documentation 2023.)*

With GameObjects explained, the Unity Editor is setup with eight key windows that are used in development to start with, with the ability to add or remove them as the user sees fit to their needs. Those eight key windows are:

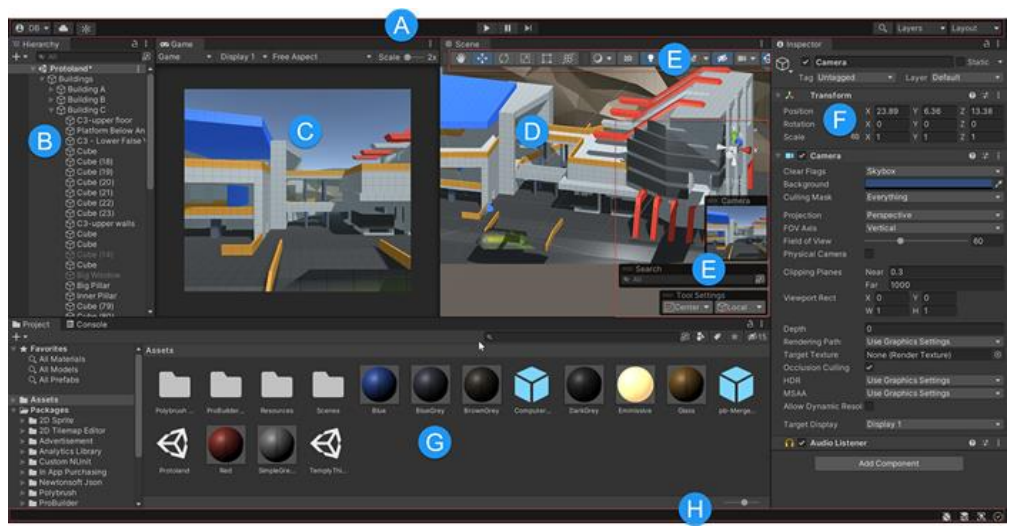

*Figure 6 – Unity Editor Window (Unity Documentation 2023.)*

1. The Toolbar, where the developer has access to their Unity Account, Unity Cloud Services, The Play Mode controls, Unity Search, The Editor Layout menu, and lastly the Undo History and Visibility Menu (the top bar of the screen, the letter A in Figure 6)
2. The Hierarchy View, where the developer can add, remove, rename, re-organise every GameObject within the game itself (in the top left, the letter B in Figure 6)
3. The Game View, which simulates what the players will see in real time and allows developers to even make edits whilst the game is emulated within the Game View window itself (in the second tab in the centre of the screen, the letter C in Figure 6)
4. The Scene View, where the main skeleton of the game is created, loaded, and worked on. This is not what the player will see once the game has been published (in the centre of the screen, the letter D in Figure 6)
5. Overlays, which contain the basic controls for manipulating the Scene View and the GameObjects within it with the ability to add custom Overlays to improve developer's workflow (in the bottom right centre of the screen, the letter E in Figure 6)
6. The Inspector View, where every Component of the selected or highlighted GameObject in either in the Scene View or in the Hierarchy can be added, removed, or tweaked (at the right of the screen next to the Game View window, the letter F in Figure 6)
7. The Project Window, which displays the entire library of assets that are available to use in the Project (bottom third of the screen below the View windows, the letter G in Figure 6)
8. The Status Bar, which provides notifications about the various processes Unity is running such as the latest Log message that has been received from the game or if there are any errors within the project stopping it from running (bottom of the whole screen, the letter H in Figure 6)

### 1.3 Creating Script files for Unity Development

Inside Unity the main way to create script files is via the user interface in tandem with an open-source model of C# (pronounced C-sharp) called Mono. C# is a modern, general-purpose object and component-oriented programming language that runs primarily in the .NET Framework and is made by Microsoft (PluralSights 2022.). Scripting this way allows it to be supported and compatible with various platforms, with the scripts themselves setup to contain everything from the core gameplay, to handle the users' inputs, to creating events based on location/input/or even time. Via the user interface, the developer can more easily alter, test and prototype different values for different parameters as well as allowing those who aren't as programming savvy to be able to understand what changes will lead to what outcome.

C# supports the classic primitive types of variables, such as int and doubles, it also supports user-defined reference types and values, and lastly all the generic methods and types which lead in general to a safer and more performance friendly experience for developers. Unlike most other assets within Unity however, scripts are usually created within Unity directly. To create a new script, you can do from either the Create menu in the top left of the Project Panel or by selecting *Assets > Create > C# Script*. The new script will be created in whichever folder was selected inside the Project panel. (Unity Documentation 2023.)

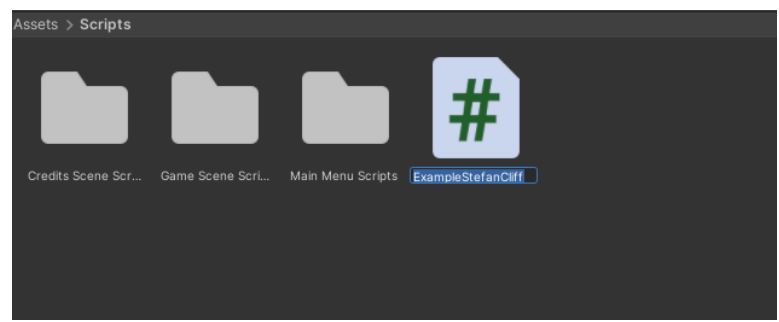

*Figure 7 – Creating and naming a new C# Script inside the Project panel*

Once the name of the new script is confirmed, by double click the file and it will open the code in a text editor which by default is Visual Studio or, in my case, Visual Studio Code. Each script is loaded up with a template by default which you can freely add to and alter as needed for your game:

*Figure 8 – Empty Template of a new Unity Script in C#*

After creating and loading our new script file, additional functionality can be added as needed to the project. However, before pressing play, first need to attach our newly created script as a Component to a GameObject within our Hierarchy Panel or via the Inspector window on a selected GameObject within our Scene View window. You can do so by either pressing the add component button at the bottom of the Inspector window, or by dragging and dropping our script file onto the GameObject/Component Inspector window itself.

*Figure 9 - Basic Example of a Unity Script in C#*

## 1.4 Tilemaps in Unity

Unity has a special and powerful system for quickly storing and handling Tile Assets for creating 2D levels, to make it easier for developers to create and iterate level design within Unity itself. The Tilemap system transfers all the required information from the Tiles placed on it to the other related components, such as a Collider which will allow the player or other colliders to interact with it. By default, the Tilemap package isn't loaded int the Unity Editor and requires the user to download it from the Package Manager, which I discuss later in this paper. When you create a Tilemap, a grid is automatically attached to the Tilemap and serves as a guide for both distance and scale. To create, change or select any tiles to be painted onto the Tilemap you use the Tile Palette which, if not already opened automatically, is in *Window > 2D > Tile Palette*.

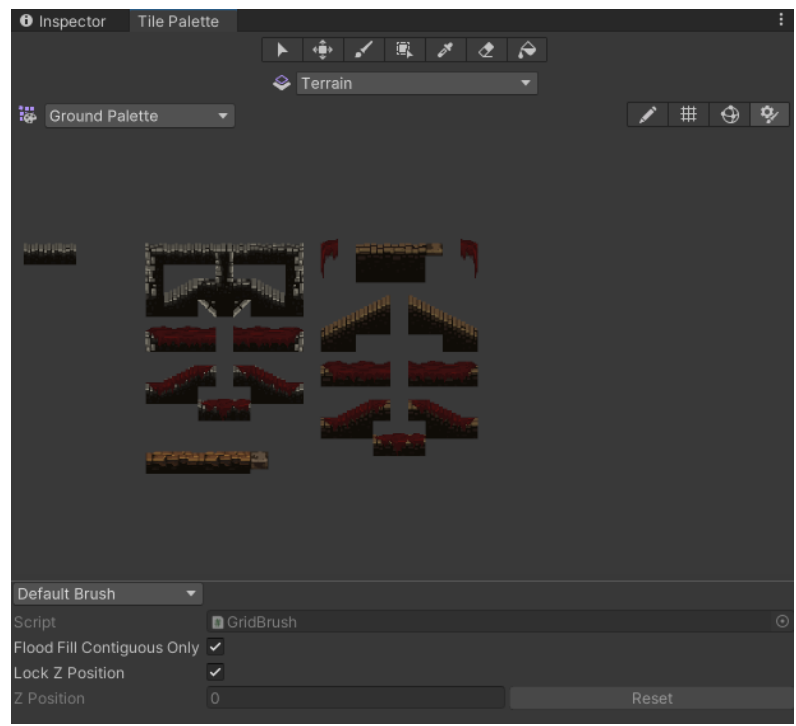

*Figure 10 – Tile Palette in Unity*

### 1.5 Assets

An asset is the representations of any single item that can be used in a game or app inside Unity. They can be visual or audio elements such as 3D models, textures, sprites, music, sound effect and more. They can also be more abstract items such as color gradients, animation mask and more. An asset might come from a file that is made by something other than Unity, such as the popular 3D modelling application Blender. (Unity Documentation 2023.) Each asset shares the same workflow:

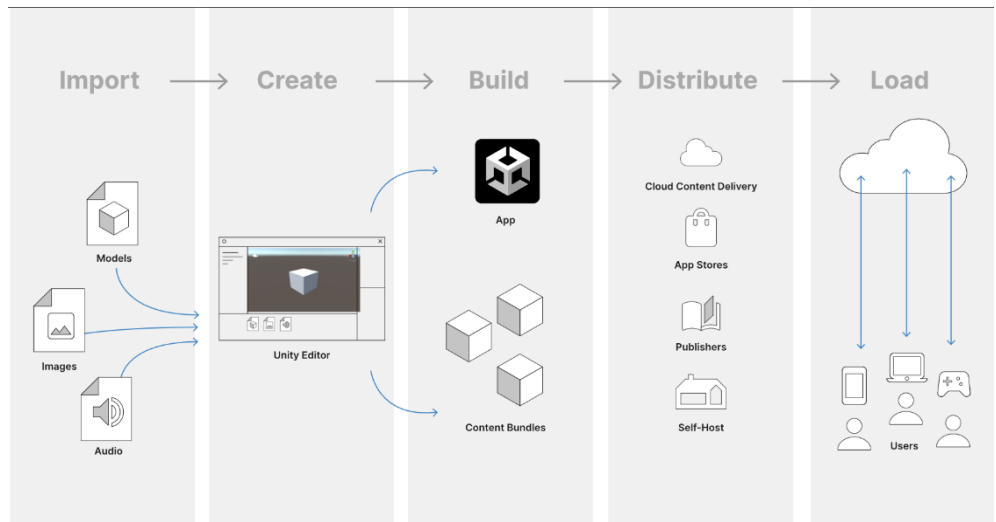

*Figure 11 – Asset Workflow in Unity (Unity Documentation 2023.)*

The above diagram shows what each asset usually goes though in a workflow, where each step means:

1. Import, loading the assets into the Unity Editor to work with
2. Create, content using the Unity Editor with those assets that are imported
3. Build, your game or app with the assets by exporting them as a binary file they can run
4. Distribute, the built product so that your users can access then
5. Load, further updates necessary at runtime

There are many types of asset files available both for free and for a price on Unity's very own Assets Store, but the two that I used the most were 2D assets and Audio. 2D Assets include sprites, textures, fonts, UI elements and more, which were crucial for the [1]development of my game. Audio assets can come in a few different extension types, the most common being *.mp3*. Inside the Assets Store after competing your purchase and acquirement of the assets you want or will need; they are linked to your Unity Account.

### 1.5.1 Package Manager

Then inside the Unity Editor under the *Window > Package Manager* you can see all the asset packs you have as well as the option to download or import them into your current

---
[1] A list of all the assets I used, as well as where I used them is available in the Appendix, with links and author credit.

- 16 -

project if they are already downloaded locally.

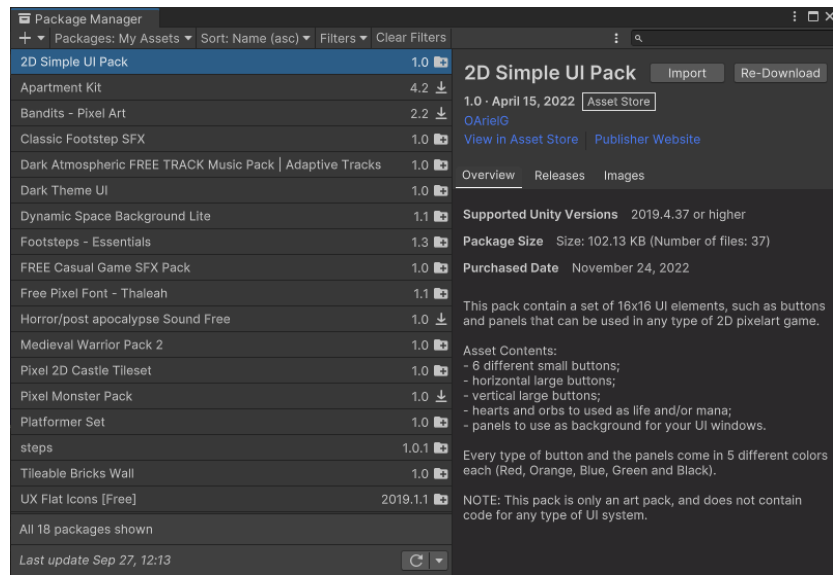

*Figure 12 – Package Manager in Unity*

## 2 Game Design, Theory and Development

This is a narrative driven 2D (two-dimensional) platformer game, where the main character is played and viewed from a distant orthographic camera that follows the player avatar. It is set in a, firstly, seemingly standard cell at the bottom of a dungeon tower with the cell door wide open and only a bed inside. The character wakes up and begins questioning who and where they are, with a sudden need to know that they set their sights on escaping in search for answers. As the player progresses throughout the level, learning how to jump and to avoid the dangers that litter the tower, a mysterious entity speaks to our main character soon after they discover that they are running out of time to escape. In a rush for answers, with time ticking down and the layout of the tower growing in scope and complexity, sooner than later they either run out of time or fall prey to the traps that creep in the crevices of the old tiles. As darkness takes them, they suddenly open their eyes once more, back at the bottom of the tower once again, seemingly in a loop.

The game development can is broken into the scenes or levels that the player interacts with, those scenes being the Main Menu, the Games Scene itself and The Credits Scene. Each scene has its own corresponding script files and designs that link and fit one another. The actual gameplay loop of the Game Scene can be broken down further into the following groups:

1. The Game Session tracker,
2. The Dialogue Scripts,
3. The Level Scripts,
4. The Player Scripts & lastly
5. The UI Scripts

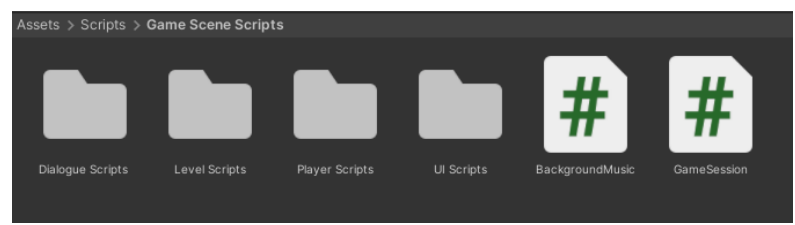

*Figure 13 – Game Scene Scripts Organisation*

The organisation of the Game Scene can be broken down into four parts, those parts being:
1. The Game Session Manager, which is an Empty GameObject onto which the Game Session.cs script file is attached
2. The Game Manager, where all the other controls and manager GameObjects are located as well as the Dialogue Manager with its DialogueTrigger groups that are turned on or off based on a variable inside of the Game Session.cs file.
3. The UI, which displays the pause button in the top right corner, informs the player of the remaining time through a blue bar that slowly empties and as well as the players current health. When the player character enters a DialogueTrigger it will display on the UI the dialogue window which shows the text, the speaker as well as a button to show the sentence.
4. The Level, which holds all the Moving Platforms, whose pathing is based on an array of waypoints set up within the Scene View, the Hazards on the level, the Background images used for the distant background that is more visible the higher the character climbs. Lastly, it holds the Tile maps of the game, those being the Terrain, the Background and The Enviroment both 1 and 2.

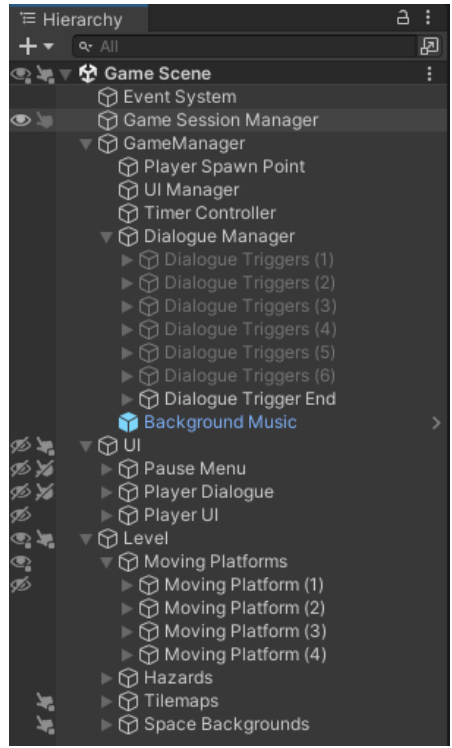

*Figure 14 – Hierarchy of The Game Scene*

## 2.1 Game Session

One core part that makes this whole game work is to have a single instance of a variable that is not destroyed once the game is booted up for the first time, and to achieve this the use of a Singleton pattern (which means that there can only be one instance of it at a time, no matter how many times it is called to create itself again) style variable called *CurrentAttempt* which always starts at 1 and increases later thanks to its NextAttempt() method that other scripts and components call to once either the player runs out of time or health. This variable is later used to control what dialogue triggers are enabled and disabled via the DialogueManager.cs, as well as controlling other logic within the game itself.

```csharp
0 references | Unity Message
private void Awake()
{
    currentAttempt = 1;

    if (instance == null)
    {
        instance = this;
        DontDestroyOnLoad(gameObject);
    }
    else
    {
        Destroy(gameObject);
        return;
    }

    if(currentAttempt == 0)
    {
        currentAttempt = 1;
    }

    dialogueManager = GetComponent<DialogueManager>();
}

2 references
public void NextAttempt()
{
    currentAttempt++;
}

1 reference
public int GetCurrentAttempt()
{
    return currentAttempt;
}
```

*Figure 15 – Core code of the GameSession.cs script file*

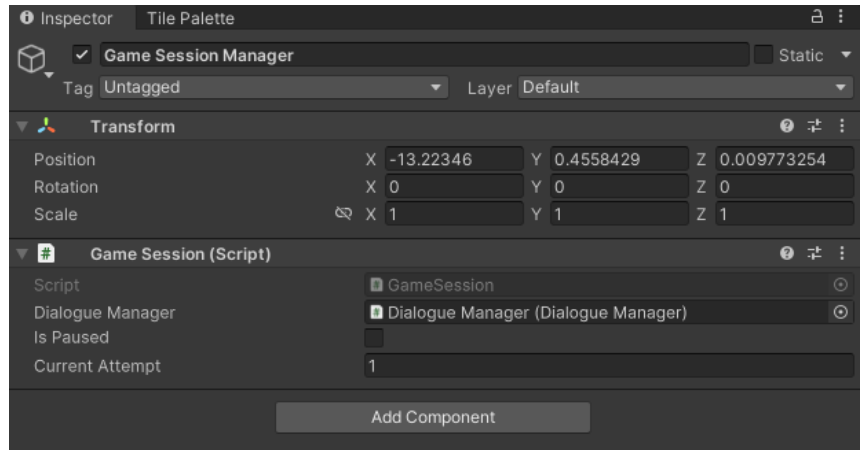
*Figure 16 – Game Session Manager in the Inspector*

## 2.2 Dialogue Scripts

To create a sense of narrative and intrigue, as well as limiting the interactions the player can have to be with the NPC be through dialogue alone, the need for an easy way to enter text into the game via the Inspector window was needed. With the class called Dialogue which would only have two variables, the speaker and an array of strings that would serve as the sentences, that speed-up in development was achieved.

```
You, 6 days ago | 1 author (You)
using System.Collections;
using System.Collections.Generic;
using UnityEngine;

You, 6 days ago | 1 author (You)
[System.Serializable]
5 references
public class Dialogue
{
    1 reference
    public string speaker;
    [TextArea (3, 10)]
    1 reference
    public string[] sentences;
}
```
*Figure 17 – Dialogue.cs Class Code*

This Dialogue class would later be used inside the DialogueTrigger.cs scripts, which would later be called in the DialogueManager.cs script file.

```csharp
0 references | You, 6 days ago | 1 author (You)
public class DialogueTrigger : MonoBehaviour
{
    1 reference
    public Dialogue dialogue;

    0 references | Unity Message
    public void OnTriggerEnter2D(Collider2D other)
    {
        if(other.gameObject.CompareTag("Player"))
        {
            DialogueManager dialogueManager = FindAnyObjectByType<DialogueManager>();
            dialogueManager.StartConversation(dialogue);

            // Here I am deactivating the trigger so that it won't run again if the player enters
            gameObject.SetActive(false);
        }
    }
}
```

*Figure 18 – The code of the DialogueTrigger.cs script file*

This script file works and is attached to empty, as in not visible, GameObjects which have a Collider that, upon anything getting in contact with it, checks to see if the object that it collided with has the tag of *Player*, if it does then it searches and calls upon the DialogueManager GameObject and it sends the dialogue information it had alongside telling the Manager to run its StartConversation() method. After making that call, it is deactivated as to not trigger the same dialogue twice. There are two altered versions of this script and these GameObjects that either start or stop the level timer, however the core functionality remains the same.

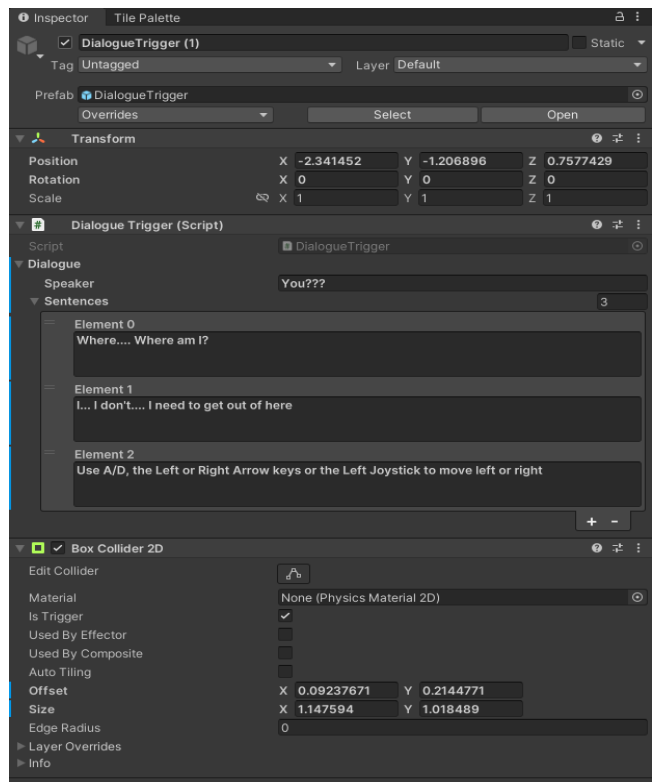

*Figure 19 – Dialogue Trigger GameObject in the Inspector*

This trigger is then put into an array inside the Dialogue Manager GameObject which holds and manages the DialogueManager.cs script file Component.

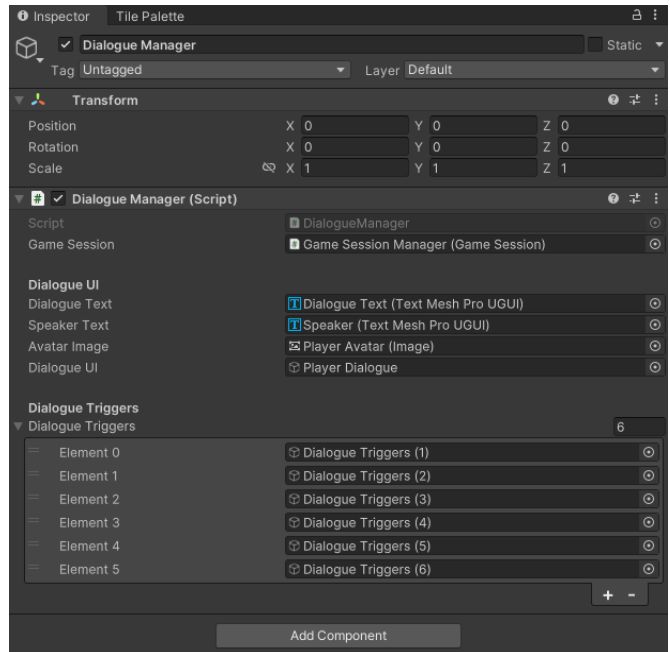

*Figure 20 – Dialogue Manager Inspector View*

This component is the one that controls and links both the Dialogue Triggers and the UI that handles the dialogue windows during the game, making sure that based on the attempt that the correct set of Dialogue Triggers are activated.

```csharp
4 references
public void StartConversation(Dialogue dialogue)
{
    Time.timeScale              = 0f;
    GameSession.instance.isPaused = true;
    speakerText.text            = dialogue.speaker;

    dialogueUI.SetActive(true);
    sentences.Clear();

    foreach (string sentence in dialogue.sentences)
    {
        sentences.Enqueue(sentence);
        //Debug.Log("Added sentence: " + sentence);
    }

    DisplayNextSentence();
}

2 references
public void DisplayNextSentence()
{
    if(sentences.Count == 0)
    {
        EndDialogue();
        return;
    }
    string sentence = sentences.Dequeue();
    StopAllCoroutines();
    StartCoroutine(TypeText(sentence));
}
```

*Figure 21 – One of three core code snippets of the DialogueManager.cs script file*

The method that the Dialogue Trigger calls, called StartConversation(), besides enabling the UI elements to be visible, it also pauses the game time to allow the user enough time to read and proceed to the next sentence. Each line is displayed with a typewriter effect, meaning that each letter appears one after the other, as to help the more organic feeling of the interaction.

```csharp
IEnumerator TypeText(string sentence)
{
    isTyping      = true;
    dialogueText.text = "";

    foreach (char c in sentence.ToCharArray())
    {
        dialogueText.text += c;

        yield return null;
    }

    isTyping = false;

}

private void EndDialogue()
{
    dialogueUI.SetActive(false);
    GameSession.instance.isPaused  = false;
    Time.timeScale                 = 1f;
}
```

*Figure 22 – The second core code snippet of the DialogueManager.cs script file*

Here lies the definition and style of execution for the typewriter effect that is achievable thanks to the special type of method called a Coroutine. A Coroutine is a special type of method that executes over multiple frames and is usually used in situations where it is needed to specify a series of events in a specific order to play out, or in a procedural animation of some kind. (Unity Documentation 2023.). They are called by using the StartCoroutine(name of the method to call as well as any defined variables it needs) method. Further on, the EndDialogue() method disables the previously enabled UI as well as unpausing the game.

```csharp
1 reference
public void UpdateDialogueTriggers(int currentAttempt)
{
    switch(currentAttempt)
    {
        case 1:
            EnableOrDisableDialogueTriggers(currentAttempt);
            break;
        case 2:
            EnableOrDisableDialogueTriggers(currentAttempt);
            break;
        case 3:
            EnableOrDisableDialogueTriggers(currentAttempt);
            break;
        case 4:
            EnableOrDisableDialogueTriggers(currentAttempt);
            break;
        case 5:
            EnableOrDisableDialogueTriggers(currentAttempt);
            break;
        case 6:
            EnableOrDisableDialogueTriggers(currentAttempt);
            break;
    }
}

6 references
public void EnableOrDisableDialogueTriggers(int attemptNumber)
{
    for (int i = 0; i < dialogueTriggers.Length; i++)
    {
        if (i == currentAttempt - 1)
        {
            dialogueTriggers[i].SetActive(true);
        }
        else
        {
            dialogueTriggers[i].SetActive(false);
        }
    }
}
```

*Figure 23 – The last third of the core logic inside the DialogueManager.cs script file*

Inside these last two methods is the core logic that lets the Dialogue Manager GameObject to enable or disable the groups of Dialogue Triggers based on the currentAttempt variable which it gets from the Game Session Manager GameObject. The only set of Dialogue Triggers that are always enable are the ones at the very end of the level, as to ensure that in case the player is either extremely skilful or somehow fails even during the sixth attempt, they can still trigger the end of the game.

```csharp
0 references | Unity Message
public void OnTriggerEnter2D(Collider2D other)
{
    if(other.gameObject.CompareTag("Player"))
    {
        DialogueManager dialogueManager = FindAnyObjectByType<DialogueManager>();
        dialogueManager.StartConversation(dialogue);
        timerController.EndGameTimer();
        if(cameraScript != null)
        {
            cameraScript.isAtEndOfLevel = true;
        }
    }
    StartCoroutine(LoadCreditsSceneAfterDelay());
}

1 reference
IEnumerator LoadCreditsSceneAfterDelay()
{
    yield return new WaitForSeconds(2f);
    SceneManager.LoadScene("Credits Scene");
}
```

*Figure 24 – The code of the EndGameScript.cs script file that are attached to the triggers at the end of the level*

Like the DialogueTrigger.cs scripts, however the EndGameScript.cs holds the addition of resetting the timer as well as pausing it alongside starting a Coroutine that after a delay of 2 (two) seconds loads the Credits Scene and effectively ending the game.

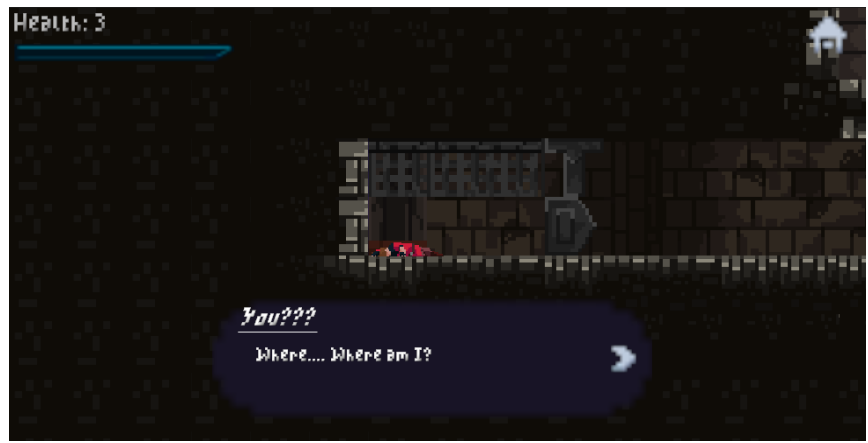

*Figure 25 – The Start of the Game, with the Dialogue Trigger showing the first sentence*

### 2.3 Narrative Design, Level Design and Scripts

The elements this game touches on, being lost, confused, relying only on what you see in front of you, questioning purpose and seeking meaning to even trying again after multiple failures. This is something many people can empathise with, especially those who use video games as a form of escapism, which has a dualistic nature when it comes to gaming (Umer Hussain, Sami Jabarkhail, George B. Cunningham, Jean A. Madsen 2021.). Often people who suffer from depression, anxiety, social issues, and such seek to escape their reality and immerse themselves in another where their issues aren't affecting them can have both a positive and negative outcome. This is often very true for people who are still in their development year where their idea of themselves is still being formed and seek to explore what and who they are by experiencing situations that would otherwise not be possible or highly unlikely. These experiences can often lead to awakenings and feeling a deeper connection or discovery, be it the discovery that what they are feeling isn't entierly only happening to them, or perhaps a realisation to do with their gender or sexuality, all of these and more can lead to possibly vast corners of their ideas of themselves (Charles Ecenbarger II 2014.).

The level design for this game was core in driving the narrative forward as well as still posing a challenge to anyone playing it, but still had to support the multiple style of players that are out there from the extremely new to the hardcore veteran players. A great important aspect this was to have a nice mixture of static and semi-dynamic elements to the level as well as obvious hazards. The multiple choice of paths to try and climb are intentional and for most of the level supports the idea of going back and forth between the left and right side, as one side always has a more static design whilst the other side will either have a moving platform or a series of more challenging jumps. At any point the player can switch over but still must be careful to not waste time transferring and to not miss their landing, as there are places where a lot of progress can be lost and effectively *lose* in that attempt to climb up within the time frame.

A part of the design for the level was to include the idea that the further the character climbs up, the more cracks in the stonework are visible, giving way to a series of colourful elements almost leaking through these seams. At the start of the level there is a larger amount of detail and almost logical thinking involved, but the higher the player character climbs, the jumps become more disconnected and are incorporating a more chaotic design as well as being more difficult and less forgiving if you miss a step. These level based changes in difficulty are an integral part of the narrative alongside the dialogue that is shared between the player character and the entity.

Splitting the tower into two sides with a series of crossover-able sections and platforms was meant to represent a series of ideas. The first time the player reaches the point where there are two ways to go, they have no knowledge beforehand and must make a quick choice, left or right. However already on the second attempt the player is now armed with some foreknowledge and might decide to stick with the known or try to risk it on the unknown side. The further the player plays, even the player character will make a comment towards the start of an attempt saying that they are growing more familiar with the tower than they are themselves.

In the end however, both paths lead to a single meeting point and mobile platform that serves as a lift that will deliver the player high enough to make the last jump onto the ending surfaces.

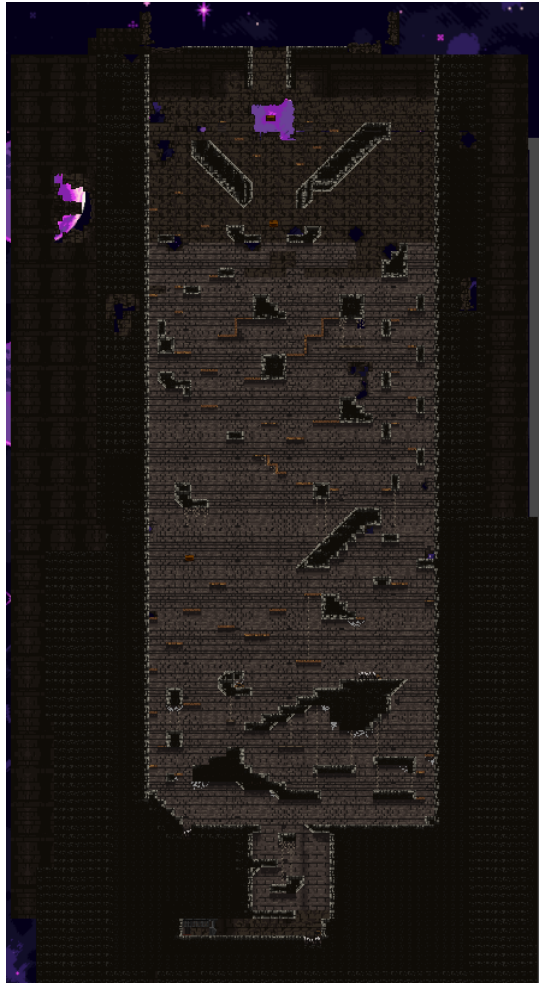

*Figure 26 – The whole level of the game*

### 2.3.1 Level Scripts

The scripts to do with level, its obstacles, and hazards as well as their interaction with the players are somewhat split between the Level Scripts folder and as well as the Player Scripts folder, when referencing certain methods such as the TakeDamage() method in the PlayerDeath.cs script file.

```
3 references
[SerializeField] private GameObject[] Waypoints;
5 references
private int currentWaypointIndex = 0;

1 reference
[SerializeField] private float platformSpeed = 2f;

0 references | Unity Message
private void Update()
{
    if(Vector2.Distance(Waypoints[currentWaypointIndex].transform.position, transform.position) < 0.1f)
    {
        currentWaypointIndex++;

        if(currentWaypointIndex >= Waypoints.Length)
        {
            currentWaypointIndex = 0;
        }
    }

    transform.position = Vector2.MoveTowards(transform.position, Waypoints[currentWaypointIndex].transform.position, Time.deltaTime * platformSpeed);
}
```

*Figure 27 – Code Snippet of the WaypointFollower.cs script file*

The moving platforms are controlled by and directed with two scripts and empty GameObjects called Waypoints, which serve as points where the moving platforms will reset its calculations and move onwards to the next one based on the order of Waypoints in its array of Waypoints. Also attached to the Moving Platform GameObjects is the StickyPlatforms.cs script file that, upon the player colliding with a platform will set the player to be the child of the platform so long as the contact stays. The reason this had to be done was to allow the player to remain idle on the platform whilst it was moving without sliding off and to ensure that the camera remained smooth during the who contact. As soon as the collision is no longer detected then the Player Character GameObject is no longer a child of that Moving Platform.

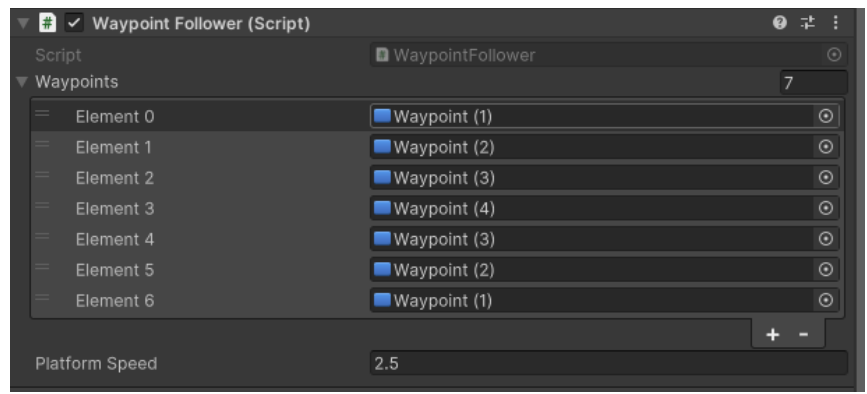

*Figure 28 –The Inspector view of the WaypointFollower.cs script file*

## 2.4 Player Character Scripts

Making sure the player character had a simple and approachable set of controls meant that more time had to be defined to focus on polish and feel. To support both keyboard and joystick input meant allowing the player to play however it is they wish and prefer. Creating a smooth and computationally do-able animation controller as well as making the jumping feel strong but the gravity not as *floaty*, meant that playtesting and finding values for the movement variables were important.

The player character is setup as a prefab, which is a system in Unity which allows me to create, configure and store a GameObject completely with all its Components, their property variables and child GameObjects all as a reusable asset. (Unity Documentation 2023.). This meant that during development, it was easier to drag and drop the player character GameObject wherever needed, which helped in testing and design for the level. By creating an empty GameObject called *Player Spawn Point* which would, once the game or attempt started, instantiate a prefab of the Player Character GameObject, as well as attach a prefab of a Camera GameObject to it to follow the Player Character.

```csharp
void Start()
{
    InstantiatedPlayerPrefab = Instantiate(PlayerPrefabToInstantiate, transform.position, transform.rotation);

    //Instantiate the camera prefab and attach the camera script to it
    GameObject cameraInstance = Instantiate(cameraPrefab, InstantiatedPlayerPrefab.transform.position, Quaternion.identity);
    cameraScript = cameraInstance.GetComponent<CameraScript>();
    cameraInstance.transform.SetParent(InstantiatedPlayerPrefab.transform);
    cameraScript.isAtEndOfLevel = false;

    if (cameraScript != null)
    {
        cameraScript.player = InstantiatedPlayerPrefab.transform; // Assign the target for the camera to follow
    }
}
```

*Figure 29 – Code Snippet of the PlaySpawn.cs script file*

This script was, as mentioned before, attached to an empty game object and it would instantiate both a Player Character prefab as well as Camera prefab on-top of it to follow the Player Characters movements. Using the Inspector window whilst highlighting the Player Spawn Point GameObject, via dragging and dropping the prefabs in their respective fields, those GameObjects would not exist till the game started. This choice was made primarily as to speed up testing and development.

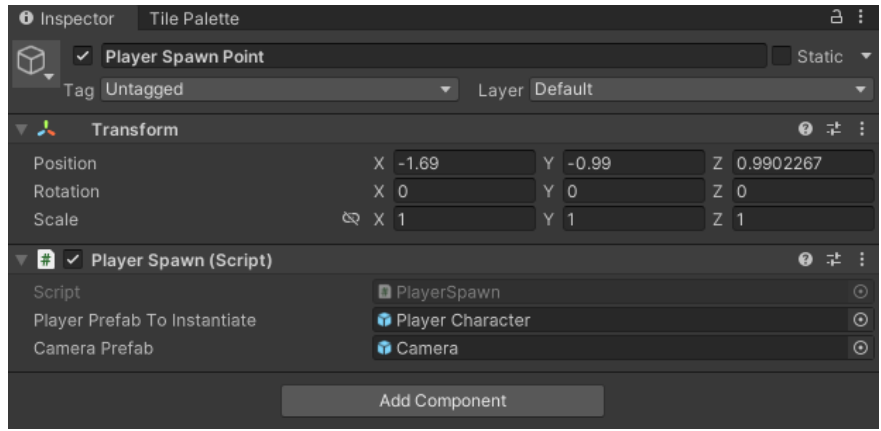

*Figure 30 – Inspector View of the Player Spawn Point GameObject*

Next came the way to handle players input as well as making sure that the Player Character cannot jump whilst already in the air. Alongside that, a way for, depending on the movement, play the correct animation and play the correct audio files.

### 2.4.1 Player Movement

```
1 reference
private void PlayerInput()
{
    if(GameSession.instance.isPaused)
    {
        return;
    }
    else
    {
        float moveInput           = Input.GetAxisRaw("Horizontal");
        bool  jumpInput           = Input.GetButtonDown("Jump");
        float joystickInputMagnitude = Mathf.Clamp(Mathf.Abs(Input.GetAxis("Horizontal")) + Mathf.Abs(Input.GetAxis("Vertical")), 0f, 1f);
```

*Figure 31 – PlayerInput() method inside the PlayerCharacter.cs script file*

Here inside the PlayerCharacter.cs script file, firstly the game checks if the game is pause, if not then it picks up any input along the horizontal axis and assigns it to a variable, where it also detects the amount the players joystick is leaning to one side or the other in case the player is using a controller.

```
4 references
private bool isGrounded()
{
    /* ...
    return Physics2D.BoxCast(playerCollider2D.bounds.center, playerCollider2D.bounds.size, 0f, UnityEngine.Vector2.down, .1f, terrainLayer);
}
...
```

*Figure 32 – The isGrounded() method inside of the PlayerCharacter.cs script file*

```
// Jumping
if (jumpInput && isGrounded())
{
    playerCharacter.velocity = new UnityEngine.Vector2(playerCharacter.velocity.x, jumpForce);
    state = MovementState.jumping;
    StopStepSound();
    jumpingSFX.Play();
}

// Falling
else if (playerCharacter.velocity.y < -0.1f)
{
    playerCharacter.velocity += UnityEngine.Vector2.up * Physics2D.gravity.y * (fallSpeed - 1) * Time.deltaTime;
    state = MovementState.falling;
    StopStepSound();
    if(isGrounded())
    {
        landingSFX.Play();
    }
}

playerAnimator.SetInteger("moveState", (int)state);
```

*Figure 33 – Jumping and increasing the falling speed of the Player Character in the PlayerCharacter.cs script file*

By creating a method that draws a second collision box that is slightly lower than that of the one surrounding the collider of the Player Character. If this newly created box is touching any part of the Tilemap that is on the terrain layer it will return true, otherwise false. This logic is used and called upon before the player wishes to jump again and will allow the code to execute only if the player is touching the ground.

The Animator's state is setup to accept the int output of an enum variable called MovementState which only has the possible following states:

```
6 references | 1 reference | 2 references | 1 reference | 1 reference
private enum MovementState { idle, running, jumping, falling } // 0 -> idle; 1 -> running; 2 -> jumping; 3 -> falling;
6 references
private MovementState state;
```

*Figure 34 – Definition of the MovementState enum inside the PlayerCharacter.cs script file*

Then with each input of the player defining the state to one of those four values, inside the Animator Controller, which is a controller that references animation clips within it and manages the various states and transitions between them by using a State Machine which could be imagined as a kind of flow-chart.

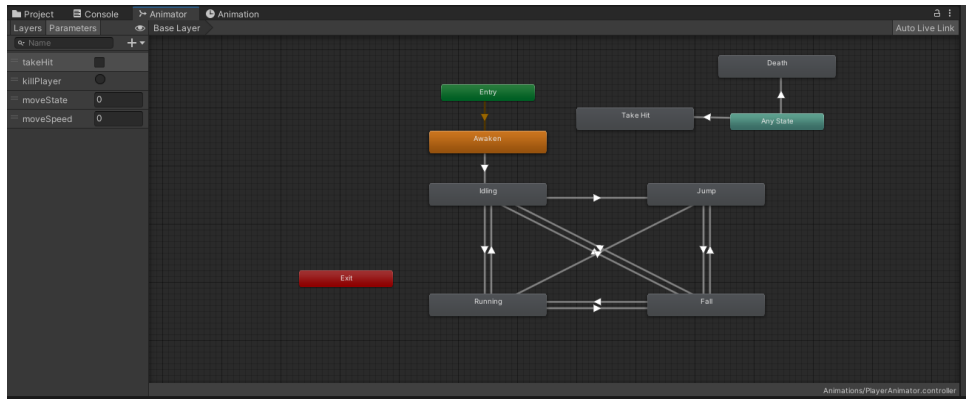

*Figure 35 – The PlayerAnimator Animation Controller which holds the relations between all the player animations*

The variables on the left are defined developer defined variables that can alter the state of our animator as needed. In my case, the switch between the four movement-based animations, control the speed of which the *Running* animation will play based on the joystick input strength to one side and two boolean values. These two boolean values are outside of the main net as they can be played based on any situation regardless of the movement state, and if they are set to true, they will:

1. takeHit will simply launch the player up a little bit akin to a bounce,
2. killPlayer will play the reverse of the Awaken animation and lie the player down whilst disabling the player's input

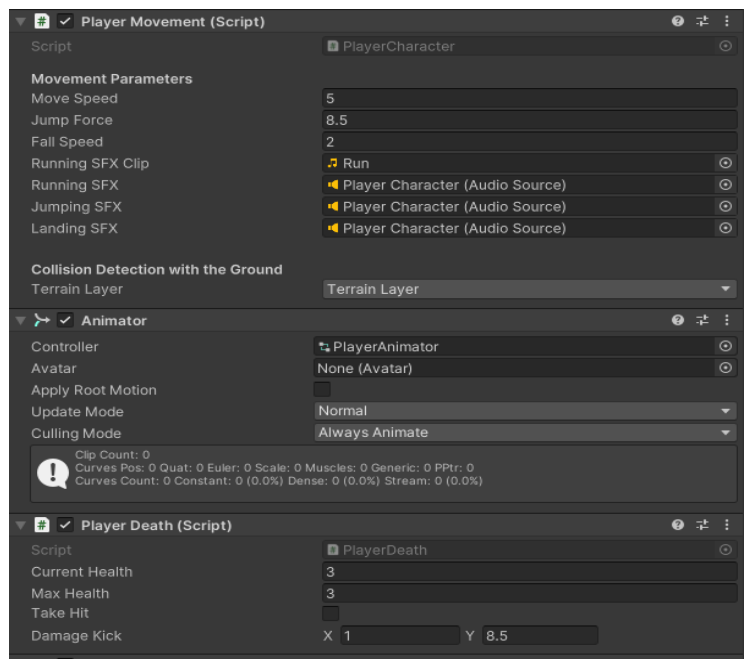

*Figure 36 – Player Character GameObject Inspector View*

### 2.4.2 Player Death

The game needed another way to restart the level other than for the timer to run out, to add a sense of caution necessary and realistic to the game for another layer of difficulty. This came in the form of spikes around the level, where if the player lands on them they will take damage. The player can only take up to three points of damage before reaching zero, thus ending the attempt and restarting the level, the same as if time ran out.

```csharp
1 reference
private void TakeDamage(int damage)
{
    takeHit = true;
    PlayerAnimator.SetBool("takeHit", takeHit);

    CurrentHealth -= damage;
    CurrentHealth = Mathf.Clamp(CurrentHealth, 0, maxHealth);

    PlayerAvatar.enabled = false;
    PlayerAvatar.enabled = true;
    PlayerCharacter.velocity = damageKick;

    takeHit = false;
    PlayerAnimator.SetBool("takeHit", takeHit);
    PlayerAvatar.enabled = true;

    if (CurrentHealth <= 0)
    {
        PlayerAnimator.SetTrigger("killPlayer");
        PlayerCharacter.bodyType = RigidbodyType2D.Static;
        RestartLevel();
    }
}
```

*Figure 37 – The TakeDamage() method inside the PlayerDeath.cs script file*

If the player character dies then the method RestartLevel() will restart the scene anew whilst also increasing the currentAttempt variable by one.

```csharp
1 reference
private void RestartLevel()
{
    GameSession.instance.NextAttempt();
    SceneManager.LoadScene(SceneManager.GetActiveScene().name);
}
```

*Figure 38 – The RestartLevel() method inside the PlayerDeath.cs script file*

```csharp
0 references | Unity Message
private void OnCollisionEnter2D(Collision2D collision)
{
    if (collision.gameObject.CompareTag("Hazard"))
    {
        TakeDamage(1);
    }
}
```

*Figure 39 – The collision checking method in case the player falls onto spikes inside the PlayerDeath.cs script file*

- 35 -

# 3 Conclusion

The journey we took developing this game, both when coming up with the idea and throughout the development and testing of it, has been rather eye opening and sparked many ideas and multiple long conversations with everyone we were able to get to test it. Most appreciated the simple design and controls, as well as the narrative that was going on both through the dialogue and throughout the level itself. We were able to get roughly 10 (ten) of our peers, people between the ages of twenty-one and twenty-nine, to try it as well a few individuals older and about 5 (five) who were younger. Our peers gave us feedback to do with the journey of it all. They felt empathic as well some knowing from personal experience what these emotions and ideas we were touching on would feel like and how to translate that feeling into a game better. Those who were older than us offered guidance to do with the Entity character, how to translate the idea of them knowing and wanting the Player Character to find the answers for themselves without giving it away. The younger demographic wanted more of a challenge and for the original time that was planned of a minute and a half to be lowered to only one minute, saying that they preferred the intensity they felt as they had to make the more difficult jumps with less chance of a second chance.

The ideas and knowledge we had going into this project be it on game development, mental health, and the representation of it, we felt would be enough to fuel us from start to finish. But, of course, that isn't the case at all. Instead, we focused on the message we wanted to send, the information and feelings we want to share with people as well as making sure the game itself is fun. Learning so much to do with programming, organisation, logic, animations, the mathematics involved in programming and so on. We learned more to do with creative writing, dialogues, level, and narrative design and how they blend.

Where we think we could work more on this game is the assets themselves, a redo of the dialogue system that we have in place of one that would likely be smoother with better controls as well as the inclusion of support for mobile phone input. The reason why is due to the little computational power and currently low graphical power needed to render the effects and images. Further work on the level itself, perhaps making it longer with the inclusion of both more hazards such as mobile buzz saws, like older platformers, as well as creating more dynamic parts for the player to interact with such as swinging platforms, climbable ropes swings, ladders, shortcuts, and the ability for keyboard players to better control how quickly or slowly they are moving.

The development of indie psychological platformer games is not terribly popular in the general media, but we believe that the market for it will only grow more and more as time goes on, as will the competition within that market as people will more likely than not seek more single-player games as they get older and older. As the technology advances so too will the quality and the quantity of new indie developers as well as developers in larger companies. New tools but also new horizons are being revealed each day, with things like Raytracing, Virtual Reality, Augmented Reality are becoming new mediums to play on mobile phones, desktops and consoles as they are all growing into more powerful devices.

12. Unity (2023.). Unity Documentation – Unity Manual – Tilemaps (version 2022.3) https://docs.unity3d.com/Manual/Tilemap.html

13. Unity (2023.). Unity Documentation – Unity Manual – Asset Workflow (version 2022.3) https://docs.unity3d.com/Manual/AssetWorkflow.html

14. Unity (2023.). Unity Documentation – Unity Manual – Coroutines (version 2022.3) https://docs.unity3d.com/Manual/Coroutines.html

15. Umer Hussain, Sami Jabarkhail, George B. Cunningham, Jean A. Madsen (2021.). The Dual nature of escapism in video gaming: A meta-analytic approach https://www.sciencedirect.com/science/article/pii/S2451958821000294

16. Charles Ecenbarger II (2014). The Impact of Video games on Identity Construction. Ball State University. https://scholar.google.com/citations?view_op=view_citation&hl=en&user=a4DGvrIAAAAJ&citation_for_view=a4DGvrIAAAAJ:d1gkVwhDpl0C

17. Unity (2023.). Unity Documentation – Unity Manual – Prefabs (version 2022.3) https://docs.unity3d.com/Manual/Prefabs.html

18. Conor Burke (2021.). Open Library - User Interface https://ecampusontario.pressbooks.pub/gamedesigndevelopmenttextbook/chapter/user-interface/#:~:text=A%20user%20interface%20in%20games,Heads%20Up%20Display%20(HUD).
- 39 -

# 5 Appendix

The public GitHub repo link of the project is available here.

## 5.1 Assets Used & Their Creators

### 5.1.1 Universal

| Item | Artist | Source/Link |
|---|---|---|
| **The MinimalPixel Font** | By Mounir Tohami | https://mounirtohami.itch.io/minimalpixel-font |
| **The GUI Sprites** | By Mounir Tohami | https://mounirtohami.itch.io/pixel-art-gui-elements |
| **Footsteps Essentials** | By Nox_Sound | https://assetstore.unity.com/packages/audio/sound-fx/foley/footsteps-essentials-189879 |
| **Dark Atmosphere FREE TRACK Music Pack \| Adaptive Tracks** | By Composed Immersion | https://assetstore.unity.com/packages/audio/music/electronic/dark-atmospheric-free-track-music-pack-adaptive-tracks-244634 |

### 5.1.2 Main Menu

| Item | Artist | Source/Link |
|---|---|---|
| **The Space Background** | By PxSprite on ArtStation.com | https://www.artstation.com/artwork/5B51aW |

### 5.1.3 Game

| Item | Artist | Source/Link |
|---|---|---|

| The Generative Space Background | By Deep-Fold | https://deep-fold.itch.io/space-background-generator |
|---|---|---|
| Platformer Set | By Szadi Art. | https://assetstore.unity.com/packages/2d/environments/platformer-set-150023 |
| Pixel 2D Castle Tileset | By Szadi Art. | https://assetstore.unity.com/packages/2d/textures-materials/tiles/pixel-2d-castle-tileset-135397 |
| Super Platformer Set | By FoxFin | https://assetstore.unity.com/packages/2d/environments/super-platformer-assets-42013 |
| Medieval Warrior Pack 2 | By Luis Melo | https://assetstore.unity.com/packages/2d/characters/medieval-warrior-pack-2-174788 |

## 5.2 Narrative Manuscript

**Attempt One**

1: Player Character: "Where… Where am I?", "I... I don't.... I need to get out of here".

2: Player Character: "I need to climb out of here...", "And I need to be careful around those spikes...", "I can likely only take a few hits before I am out for good..."

3: Player Character: "Time... I'm losing time".

4.1: Entity: "Lost, are we?"

4.2: Player Character: "Who are you?".

4.3: Entity: "Wrong question…".

**Attempt Two**

1.1: Player Character: "No... No, I must climb up!", "Why do I have to climb up?".

1.2: Entity: "Still there, are you?"

1.3: Player Character: "Where am I?".

1.4: Entity: "Wrong again…"

2.1: Player Character: "Higher... I must go higher...", "What is higher?".

2.2: Entity: "Truly, you must begin to learn to ask the right questions if you want answers.".

2.3: Player Character: "What are the right questions?".

2.4: Entity: "Wrong again...".

**Attempt Three**

1.1: Player Character: "Why am I here? If I climb, I'll know why, it must have the answers."

1.2: Entity: "How can you be so certain?"

1.3: Player Character: "Where are the answers then?"

1.4: Entity: "I'm sure by now you know what I'll say next."

1.5: Player Character: "Wrong question?"

1.6: Entity: "Quick learner, aren't we?"

**Attempt Four**

1: Player Character: "This tower... These walls... These floors." , "I feel more familiar with them than I do myself at this point.", "But I somehow feel the most lost I have since I first woke up here...", "Why am I still doing this?", "What is this feeling I know deep in my gut, but I cannot explain?", "Who is it that speaks to me, distant yet almost teasing me. Egging me onwards.", "Where am I going? How will I know I am there?"

2: Player Character: "What did I do to deserve this?"

**Attempt Five**

1: Player Character: "…"

**Attempt Six**

1.1: Player Character: "Surely, one of these attempts will work? If not, why do I still attempt them?" , "Where did that voice go? What did it even sound like again?", "...", "…", "I don't remember, maybe it was all in my head...".

1.2: Entity: "You really think that you could have imagined me up when you don't even know your own name?" , "Who the hell told you to mope around?", "I thought you were making it to the top weren't you?", Pick up the pace, pip your step and do what you wanted to do!".

  2.1: Player Character: "I'm running out of time, it isn't enough!"

  2.2: Entity: "What time are you talking about?"

  **End Scene**

  1.2: Player Character: "Wait..." , "Where is everyone?  ", "Where am I?", "Who am I?", "When am I?"

  1.2: Entity: "The correct question deserves a correct answer…", "To answer simply, you aren't... At least not in the way you understood things.", "You aren't anywhere, anyone at any point in time.", "You simply are and thusly aren't, you were and will be but also never was nor never will you be.", "There is time to explain this all over some coffee", "We might be timeless beings of the known and unknown cosmos but it doesn't mean we have to be savages.", "Do you take yours with cream and sugar?"